\documentclass[twoside]{dis08}
\usepackage[latin1]{inputenc}
\usepackage[dvips]{graphicx,epsfig,color}
\usepackage{wrapfig,rotating}
\usepackage{amssymb,amsmath,array}

\begin{document}
\title{Hard Electroproduction of Vector Mesons}

\author{Peter Kroll 
\vspace{.3cm}\\
Universit\"at Wuppertal - Fachbereich Physik\\
D-42097 Wuppertal - Germany}

\maketitle

\begin{abstract}
It is reported on a global analysis of hard vector-meson electroproduction
which is based on the handbag factorization. The generalized parton
distributions are constructed from their forward limits with the help
of double distributions and the partonic subprocesses are calculated
within the modified perturbative approach.  
\end{abstract}

\begin{wrapfigure}{r}{0.5\columnwidth}
\centerline{\includegraphics[width=0.45\columnwidth,bb=20 340 547 744]
{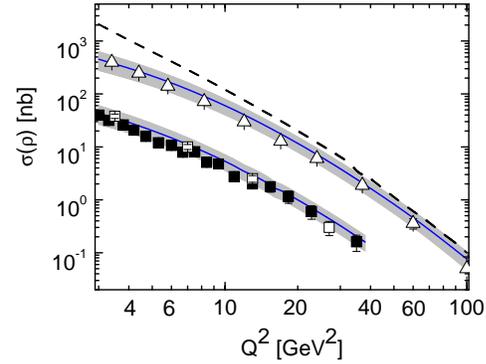}}
\caption{The cross section for $\rho^0$ production at $W=90$ and $75\,{\rm GeV}$
(the latter is divided by 10). Data taken from ZEUS \cite{zeus,zeus98}
  (open triangles and squares) and H1 \cite{h1} (solid squares). The
  shaded bands indicate the uncertainties of the theoretical results.}
\label{Fig:cross}
\end{wrapfigure}

In this talk I am going to report on an ongoing analysis
\cite{gk1,gk2,gk3} of hard electroproduction of light vector mesons
within the handbag approach which is based on QCD factorization into
hard subprocesses and generalized parton distributions (GPDs) encoding
the soft, non-perturbative physics. As is well-known the leading-twist
contribution (i.e.\ employing the collinear approximation) evaluated
to leading-order of perturbative QCD, overestimates the cross
section for the processes of interest drastically at photon
virtualities, $Q^2$,  of about $10\,{\rm GeV}^2$. This effect
diminishes with increasing $Q^2$ but the predictions to leading-twist
accuracy are still larger than experiment \cite{zeus} at, say, 
$100\,{\rm GeV}^2$, see dashed line in Fig. \ref{Fig:cross}.
As shown in Ref.\ \cite{kugler} the next-to-leading-order QCD corrections
are very large. However a recent attempt \cite{ivanov} to resum higher
orders seems to indicate that the sum of all higher order corrections
is not large. In view of this unsettled situation it seems to be
reasonable to simply use the leading order and add power corrections
to it. To do so we calculated in our previous work \cite{gk1,gk2,gk3} 
the quark ($\gamma^* q\to Vq$) and gluon ($\gamma^* g\to Vg$) subprocess
amplitudes within the modified perturbative approach \cite{botts89} in
which quark transverse degrees of freedom as well as Sudakov
suppressions are taken into account. This approach
allows to calculate not only the asymptotically dominant
(longitudinal) amplitude for $\gamma^*_Lp\to V_Lp$ but also the one
for transversely polarized photons and vector mesons ($\gamma^*_Tp\to
V_Tp$). In constrast to the longitudinal amplitude the latter one
cannot be calculated in collinear approximation since it suffers from 
infrared singularities \cite{man,teryaev}. The quark transverse momenta, 
${\bf k}_\perp$, provide an admittedly model-dependent regularization
scheme of these singularities by replacements of the  type 
\begin{equation}
    \frac1{d Q^2} \longrightarrow \frac1{d Q^2 + {\bf k}^2_\perp}\nonumber
\end{equation}
in the parton propagators. Here, $d$ is a momentum fraction or a
product of two. The kinematical region considered in \cite{gk1,gk2,gk3}
is characterized by small skewness ($\xi\leq 0.1$) and small invariant
momentum transfer ($-t\leq 0.5\,{\rm GeV}^2$) but large photon
virtuality ($Q^2\geq 3\,{\rm GeV}^2$) and large energy in the
photon-proton c.m.s.\ ($W\geq 5\,{\rm GeV}$). In this kinematical
region the dominant helicity amplitudes for the process 
$\gamma^* p\to Vp$ are given by
\begin{eqnarray}
{\cal M}^V_{\mu +,\mu +}&=&\frac{e}{2}\sum_a e_a {\cal C}_V^a \left\{\langle
H\rangle^g_{V\mu} + \langle H\rangle^a_{V\mu} + \langle \widetilde{H}\rangle^g_{V\mu} 
+ \langle \widetilde{H}\rangle^a_{V\mu}\right\}\,,   \nonumber\\
{\cal M}^V_{\mu -,\mu +}&=&-\frac{e}{2}\frac{\sqrt{-t}}{2m}\,
\sum_a e_a {\cal C}_V^a \left\{\langle
E\rangle^g_{V\mu} + \langle E\rangle^a_{V\mu}\right\}\,,
\label{amplitudes}
\end{eqnarray}
in which some simplifications, relevant for the small $\xi$ region,
have been used.  Explicit helicity labels refer to the proton while
$\mu$ denotes the helicity of the photon and the vector meson. The
quark flavors are denoted by $a$ and $e_a$ is the corresponding
charge. The non-zero flavor weight factors read for the vector mesons
of interest 
\begin{equation}
{\cal C}_\rho^{\,u} =-{\cal C}_\rho^{\,d} = {\cal C}_\omega^{\,u} 
={\cal C}_\omega^{\,d} =1/\sqrt{2}\,, \qquad {\cal C}_\phi^{\,s}= 1\,.
\label{flavor}
\end{equation}
The terms $\langle F\rangle$ denote convolutions of subprocess
amplitudes and GPDs ($F=H, E, \widetilde{H}$). Explicitly they read
($i=g,a$, $x_g=0$, $x_a=-1$)
\begin{equation}
\langle F\rangle^i_{V\mu} = \sum_\lambda \int_{x_i}^1 dx\, 
{\cal H}^{Vi}_{\mu\lambda,\mu\lambda}(x,\xi,Q^2,t=0)\,F^i(x,\xi,t)\,.
\end{equation}
The helicity of the partons participating in the subprocess, is
labelled by $\lambda$. Note that $\langle \widetilde{H}\rangle^i_{V0}$.
The subprocess amplitudes ${\cal H}$ are calculated in the impact
parameter space
\begin{equation}
{\cal H}^{Vi}_{\mu\lambda,\mu\lambda} = \int d\tau d^2b\, 
         \hat{\Psi}_{V\mu}(\tau,-{\bf b})\, 
      \hat{\cal F}^{i}_{\mu\lambda,\mu\lambda}(x,\xi,\tau, Q^2,{\bf b})\, 
         \alpha_S(\mu_R)\,{\rm exp}{[-S(\tau,{\bf b},Q^2)]}\,.
\label{mod-amp}
\end{equation}   
Its $t$ dependence is neglected for consistency since it provides
corrections of order $t/Q^2$ which are generally neglected. On the
other hand, the $t$ depencence of the GPDs is taken into account
since in them $t$ is scaled by a soft parameter. The hard scattering
kernels $\hat{\cal F}$, or their respective Fourier transform ${\cal F}$, 
are calculated to leading-order of perturbative QCD including quark
transverse momenta. The explicit expressions can be found in
\cite{gk1,gk2,gk3}. Also for the Sudakov factor $S$ in (\ref{mod-amp}), 
the renormalization ($\mu_R$) and factorization ($\mu_F$) scales it is 
refered to these articles. 

For the light-cone wave function $\Psi$
Gaussians in the quark transverse momenta are used
($\Psi \sim \exp{[-a^2_{Vj}{\bf k}^2_\perp/(\tau(1-\tau)]}$) with transverse
size parameters $a_{Vj}$ fitted to experiment. Here, $j$ refers to
either longitudinally or transversally polarized vector mesons and
$\tau$ is the momentum fraction of the quark entering the meson. 

\begin{figure}[h]
\begin{center}
\includegraphics[width=0.46\columnwidth,bb=41 350 526 728]
{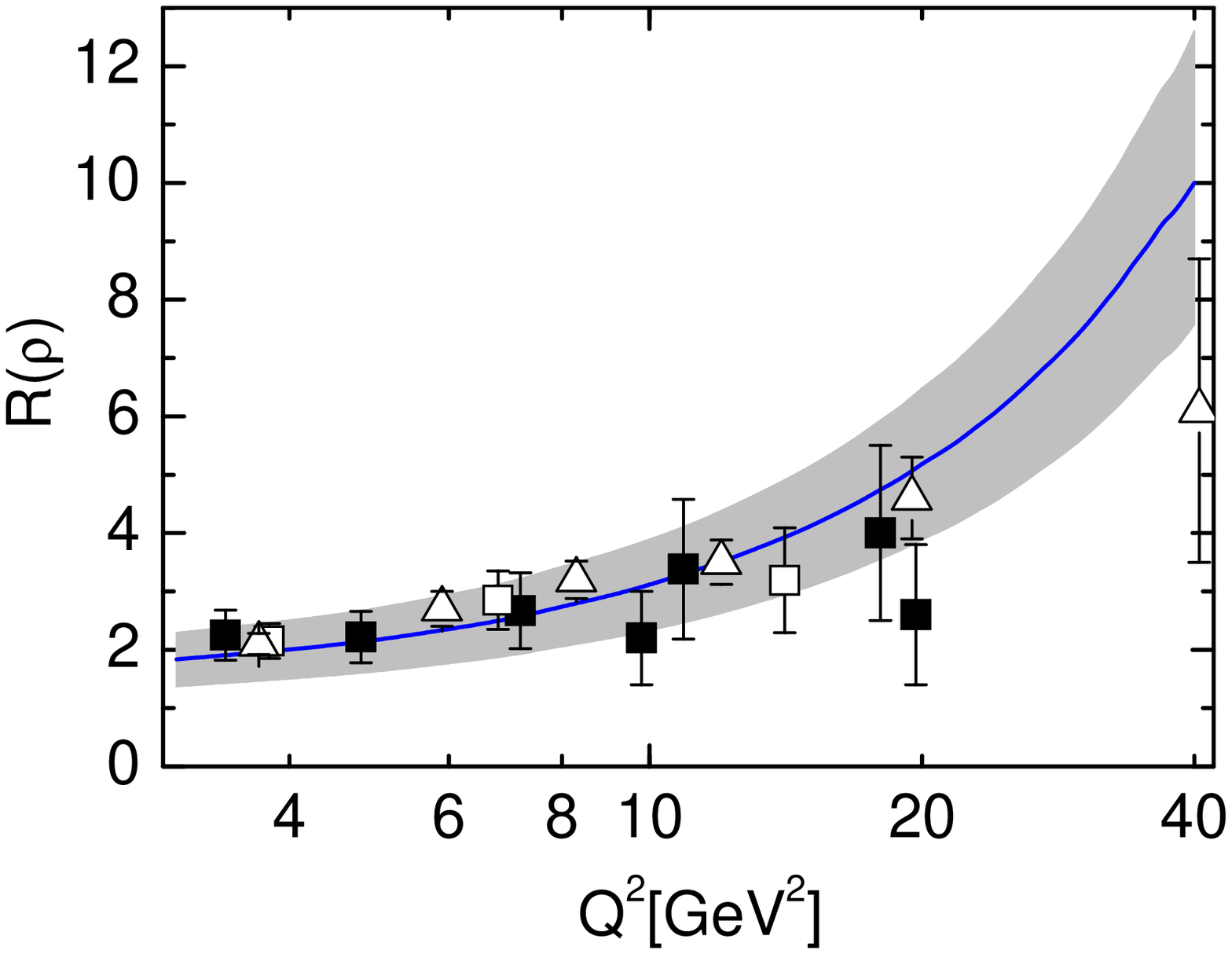}
\includegraphics[width=0.43\columnwidth,bb=53 321 540 738]
{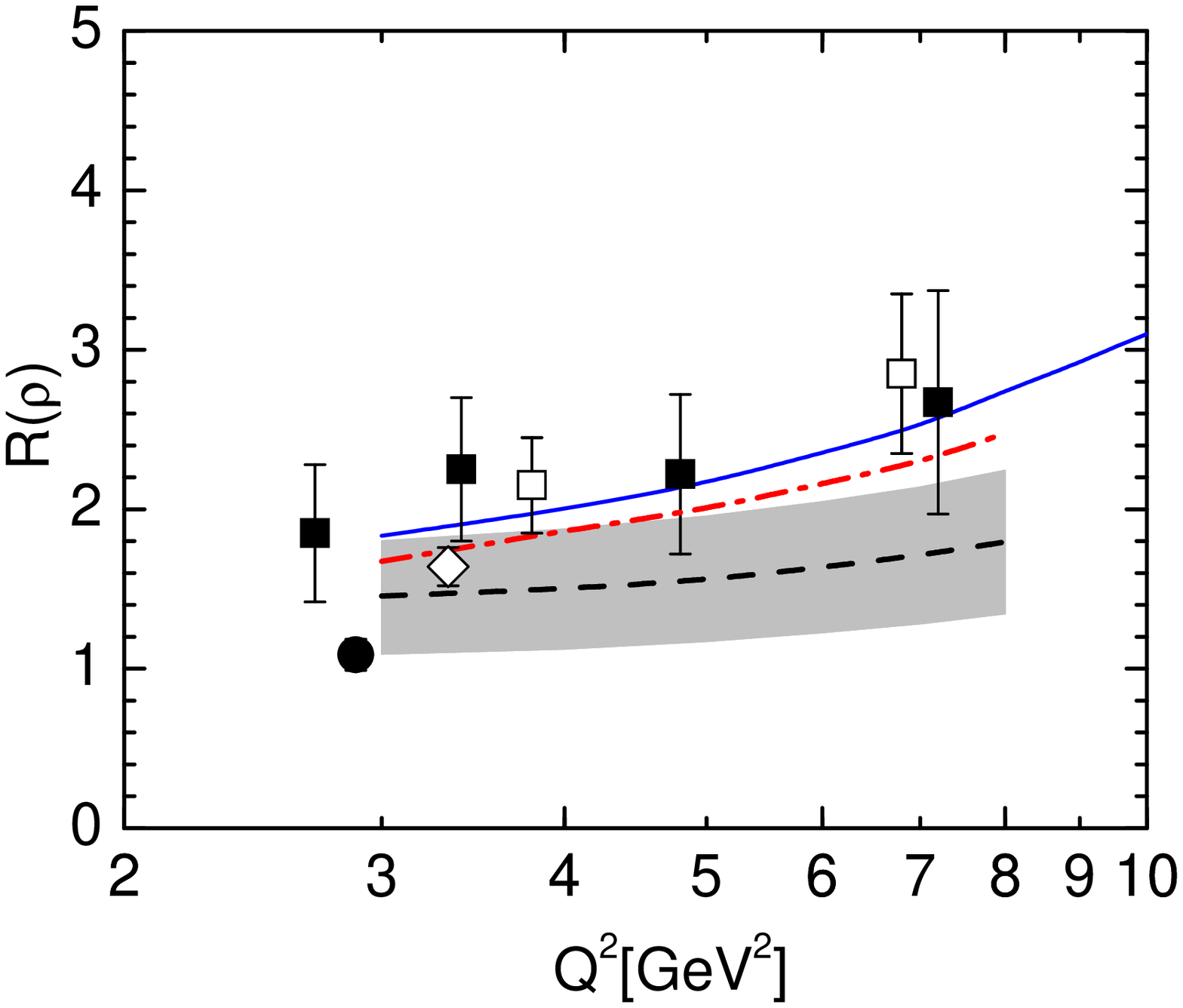}
\end{center}
\caption{The ratio of longitudinal and transversal cross section 
  for $\rho^0$ production at $W=90$ (left) and $75 (10,5)\,{\rm GeV}$ 
  (right) shown as solid (dash-dotted, dashed) lines. Preliminary data 
  from HERMES \cite{hermes-draft} and COMPASS \cite{compass-prel} are 
  shown as solid circle and diamond, respectively. For further
  notation refer to  Fig.\ \ref{Fig:cross}.}
\label{Fig:R}
\end{figure}

For the cross sections of $\gamma^*_{L(T)}p\to V_{L(T)}p$ measured
with unpolarized protons, only the GPD $H$ is to be considered, the
contributions from the other two GPDs can be neglected. The GPD $H$ is
constructed from the CTEQ6 parton distributions \cite{cteq6} using the
familiar double distribution ansatz. Figure \ref{Fig:cross} compares
the predictions for the longitudinal cross section of $\rho^0$
production with the HERA data \cite{zeus,zeus98,h1}. Satisfactory agreement
between theory and experiment is achieved. Another example of results,
namely the ratio of longitudinal and transverse cross sections
is shown in Fig. \ref{Fig:R}. Again good agreement is to be seen. For
a detailed comparison of the results with the data on $\rho^0$ and $\phi$
production from HERA, FNAL, COMPASS and HERMES can be found in \cite{gk2,gk3}.
The predicted cross section for $\phi$ production is even in agreement
with recent data from CLAS \cite{clas08} in spite of the fact that $W$
lies outside of our presupposed range of kinematics.

From the results shown in Figs.\ \ref{Fig:cross} and \ref{Fig:R}, it  is
obvious that the magnitudes of both the longitudinal and transverse
amplitudes, are correctly predicted. Their relative phase can be tested
by the spin density matrix elements ${\rm Re}\, r^5_{10}$ and ${\rm Im}\,
r^6_{10}$. It turns out that the proposed handbag approach predicts a 
relative phase of about $3^\circ$ while experiment requires a much
larger phase although with strong fluctuations ($10 - 30^\circ$). 
Whether the model for the transverse amplitude, which represents a power
correction to the leading longitudinal one, is inadequate for this
detail needs further investigation. However, that the sum 
${\rm Re}\, r_{10}^5+{\rm Im}\, r_{10}^6$ amounts to only $1\%$ of the 
corresponding difference of these SDME makes it clear that the neglected 
helicity flip $\gamma^*\to V$ transitions  are not responsible for the
observed conflict.

\begin{wrapfigure}{r}{0.5\columnwidth}
\centerline{\includegraphics[width=0.45\columnwidth,bb=10 316 535 754]
{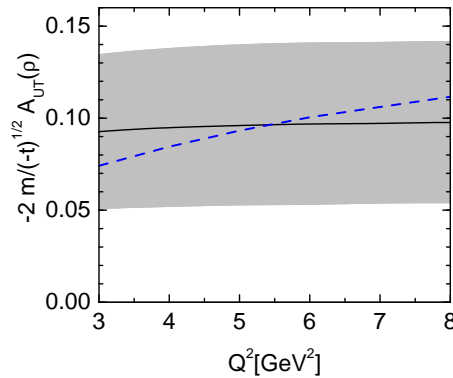}}
\caption{The asymmetry $A_{UT}$,  scaled by $-2m/\sqrt{-t}$, for $\rho^0$ 
production versus $Q^2$ at $W=5\,{\rm GeV}$ and evaluated at
$t=0$. The dashed line represents the leading-twist contribution. }
\label{Fig:AUT}
\end{wrapfigure}

The roles of the GPDs $\widetilde{H}$ and $E$ can only be explored
with polarization data where interference terms between $H$ and
$\widetilde{H}$ or $E$ are probed. Analogously to $H$ one may also
construct $\widetilde{H}$ and $E$ from their forward limits with
the double distribution ansatz. In the case of $\widetilde{H}$ the
forward limit is given by the polarized parton distributions while the 
forward limit of $E$ has been determined from the data on the Pauli 
form factor of the nucleon in an analysis of the zero-skewness GPDs 
\cite{dfjk4}. Evaluating these GPDs it turns out that both these GPDs 
are dominated by the valence quark contributions while the sea and the
gluon contributions seem to be small \cite{gk3,kugler}. This feature
is to be contrasted whith the behavior of $H$ where the gluon plays 
the most prominent role (except at very large $x$). A consequence of
these characteristics is that polarization effects like the initial state 
helicity correlation $A_{LL}$ or the target asymmetry $A_{UT}$  are 
generally small and disappear with increasing energy. Particularly
small are such observables for $\phi$ production since the valence 
quarks of the proton do not contribute to this process. Also small but 
clearly non zero effects are obtained for $\rho^0$ production. As an 
example the target asymmetry, measuring the imaginary part of the 
interference between the proton helicity flip and non-flip amplitudes 
and, hence, between $H$ and $E$ (see Eq.\ (\ref{amplitudes})) is shown 
in Fig.\ \ref{Fig:AUT}. Larger effects are found for $\omega$
production since the sum $e_u F^u_{\rm val}+e_d F^d_{\rm val}$ of the 
GPDs $\widetilde{H}$ or $E$ occurs (see Eq.\ (\ref{flavor})) and not 
the difference as for $\rho^0$ production.  Given that $F^u_{\rm val}$ 
and  $F^d_{\rm val}$ have opposite signs which follows from the known 
lowest moments of their forward limits (see the discussion in
\cite{gk3}) the mentioned sum of both is much larger than their difference. 
A preliminary HERMES result \cite{hermes-aut} for $\rho$ production, 
integrated on the range $0\leq -t \leq 0.4\,{\rm GeV}^2$, is 
$-0.033\pm 0.058$ at $Q^2=3.07\,{\rm GeV}^2$ and $W=5\,{\rm GeV}$
while a value of $-0.02\pm 0.01$ for this kinematical situation is
found in \cite{gk3}. 

In summary - the handbag approach proposed in \cite{gk1,gk2,gk3} which
consists of GPDs constructed from double distributions, Gaussian wave 
functions for the vector meson and power corrections generated from
quark transverse momenta in the subprocess describes the data on light
vector meson electroproduction measured by HERMES, COMPASS, FNAL and
HERA over a wide range of kinematics. While the GPD $H$ is well fixed
by the existing data the other GPDs are not severely constrained as
yet. More polarization data are required here. 

\begin{footnotesize}

\end{footnotesize}


\end{document}